\documentclass[10pt,conference]{IEEEtran} 
\usepackage[table]{xcolor}
\usepackage{hyperref}
\usepackage{balance}
\usepackage{tcolorbox}
\usepackage{soul}
\usepackage{listings}
\usepackage{pgfplots}
\pgfplotsset{compat=1.18} 
\usepackage{multirow}
\usepackage{booktabs} 
\usepackage{subcaption} 
\input{marcos}
\usepackage{fontawesome5}
\usepackage{fancyhdr}

\usepackage{paralist}

\usepackage{amsmath} 
\usepackage{cleveref}

\lstdefinestyle{inlinecode}{basicstyle={\ttfamily\scriptsize\bfseries}}

\newcommand{\urls}[1]{{\scriptsize\url{#1}}}
\usepackage{tcolorbox}

\usepackage{paralist}
\usepackage[outercaption]{sidecap}
\usepackage{caption}
\usepackage [autostyle, english = american]{csquotes}
\MakeOuterQuote{"}
\newcounter{scn}
\usepackage{balance}

\newif\ifpienumberinlegend
\pgfkeys{/number in legend/.code=
    \expandafter\let\expandafter\ifpienumberinlegend
    \csname if#1\endcsname
    \ifpienumberinlegend

    \def\beforenumber##1\afternumber{}%
    \fi,
    /number in legend/.default=true
}
\definecolor{1c1}{RGB}{188,162,6}
\definecolor{1c2}{RGB}{137,129,80}
\definecolor{1c3}{RGB}{239,167,31}
\definecolor{1c4}{RGB}{88,194,241}
\definecolor{1c5}{RGB}{6,180,188}

\usepackage{tcolorbox}
\tcbuselibrary{listings,skins}
\tikzset{mynode/.style={draw=white,solid,circle,fill=green,inner sep=1pt, thick,
text=black}}
\tikzset{arrow line/.style={dashed, line width= 2.5pt, color=#1}}

\usepackage{bchart}
\usepackage{xcolor}
\usepackage{framed} 
\usepackage{enumitem}

\tikzstyle{chart}=
[legend label/.style={font={\small},anchor=west,align=left},
legend box/.style={rectangle, draw, minimum size=5pt},
axis/.style={black,thin,->},
axis label/.style={anchor=east,font={\tiny}}]

\tikzstyle{bar chart}=[
chart, bar width/.code={
    \pgfmathparse{##1/2}
    \global\let\bar@w\pgfmathresult},
bar/.style={thin, draw=black},
bar label/.style={font={\bf\small},anchor=north},
bar value/.style={font={\footnotesize}},
bar width=.75,]

\tikzstyle{pie chart}=
[chart,
slice/.style={line cap=round, line join=round, thin, draw=black},
pie title/.style={font={}},
slice type/.style 2 args={
    ##1/.style={fill=##2},
    values of ##1/.style={}}]

\pgfdeclarelayer{background}
\pgfdeclarelayer{foreground}
\pgfsetlayers{background,main,foreground}

\newcommand{\tempsum}{0}

\newcommand{\pye}[3][]{
    \begin{scope}[#1]
        \pgfmathsetmacro{\curA}{0}
        \pgfmathsetmacro{\r}{1}
        \def\c{(0,0)}
        \node[pie title] at (270:1.3) {#2};
        \def\tempsum{0}
        \foreach \v/\s in{#3}{
          \pgfmathparse{\v+\tempsum}
          \global\let\tempsum=\pgfmathresult}
        \foreach \v/\s in{#3}{
            \pgfmathsetmacro{\deltaA}{\v/\tempsum*360}
            \pgfmathsetmacro{\nextA}{\curA + \deltaA}
            \pgfmathsetmacro{\midA}{(\curA+\nextA)/2+7}
            \path[slice,\s] \c
            -- +(\curA:\r)
            arc (\curA:\nextA:\r)
            -- cycle;
            \pgfmathsetmacro{\d}{min(1.4,max((2.5-0.1*\v) , .5)}

            \begin{pgfonlayer}{foreground}
                \path \c -- node[pos=\d,pie values,values of \s]{$\v\%$} +  (\midA:\r);
            \end{pgfonlayer}
            \global\let\curA\nextA}
    \end{scope}}

\newcommand{\legend}[2][]{
    \begin{scope}[#1]
        \path
        \foreach \n/\s in {#2}
            {
                  ++(0,-10pt) node[\s,legend box] {} +(3pt,0) node[legend label] {\n}
            }
        ;
    \end{scope}}
	
\def\test#1{%
    \ifnum #1 > 0
      #1
    \fi
}

\newcommand{\lstbg}[3][0pt]{{\fboxsep#1\colorbox{#2}{\strut #3}}}
\lstdefinelanguage{diff}{
  morecomment=[f][\lstbg{red!20}]-,
  morecomment=[f][\lstbg{green!20}]+,
  morecomment=[f][\textit]{@@},
}

\newcommand{\customBarChart}[3]{%
    \begin{tikzpicture}[scale=0.8]
        \begin{axis}[
            xbar stacked,   
            xmin=-50,         
            xmax=90,         
            ytick=\empty,     
            y dir=reverse,    
            bar width=2mm,    
            width=8cm,        
            height=18mm,      
            hide y axis,      
            axis x line*=middle, 
            major tick length=0pt, 
            minor tick length=0pt, 
            axis line style={draw=none}, 
            xticklabels={},
            tick style={ 
                draw=gray,
                major tick length=8pt,
                tick align=center,
            },
            extra x ticks={0}, 
            extra x tick labels={}, 
            extra x tick style={ 
                grid=major,
                grid style={black},
            },
            axis line style={gray}, 
        ]
        \addplot+[xbar, fill=yellow!90, draw=none, point meta=explicit symbolic] coordinates {(#1,0) [#1\%]};
        \addplot+[xbar, fill=red!90, draw=none, point meta=explicit symbolic] coordinates {(#2,0) [#2\%]};
        \addplot+[xbar, fill=green!70!black, draw=none, point meta=explicit symbolic] coordinates {(#3,0) [#3\%]};
        \end{axis}
    \end{tikzpicture}
}

\newcommand{\customTwoBarChart}[8]{%
\pgfmathsetmacro{\resultxmin}{#2 + #3}
\pgfmathsetmacro{\resultxminnext}{#5 + #6}
\begin{figure}[!h]
    \centering
    \hspace{-2cm}
    \begin{subfigure}[b]{.3\columnwidth}
        \centering
    \begin{tikzpicture}[scale=0.6]
        \begin{axis}[
            xbar stacked,   
            xmin=\resultxmin,         
            xmax=#1,        
            hide y axis,      
            axis x line*=bottom, 
            width=8cm,        
            height=20mm,      
            y dir=reverse,  
            legend style={at={(0.5,2.65)},anchor=north,legend columns=-1,draw=none}, 
            xticklabels={40\%,20\%,0,20\%,40\%,60\%}, 
            xticklabel style={anchor=north}, 
            tick align=center,
            bar width = 15pt,
            extra x ticks={0}, 
            extra x tick style={ 
                grid=major,
                grid style={black},
                tick label style={anchor=north},
            },
        ]
        \addplot+[xbar, fill=lightgray!90!white, draw=lightgray!90!white, mark=none,forget plot] coordinates {(#2,0)};        
        \addplot+[xbar, fill=orange!85!white, draw=orange!85!white, mark=none, forget plot] coordinates {(#3,0)};
        \addplot+[xbar, fill=cyan!95!white, draw=cyan!95!white, mark=none, forget plot] coordinates {
            (#1,0)
        };

        \addlegendimage{orange,fill=orange}
        \addlegendentry{No}
        \addlegendimage{lightgray,fill=lightgray}
        \addlegendentry{Maybe}
        \addlegendimage{cyan,fill=cyan}
        \addlegendentry{Yes}
        \end{axis}
    \end{tikzpicture} 
    \label{fig:task-interest}
    \end{subfigure}
\hspace{1.5cm}
    \begin{subfigure}[b]{.2\columnwidth}
        \centering
        \begin{tikzpicture}[scale=0.6]
            \begin{axis}[
                xbar stacked,   
                xmin=\resultxminnext,         
                xmax=#4,        
                hide y axis,      
                axis x line*=bottom, 
                width=8cm, 
                height=20mm,
                y dir=reverse,  
                legend style={at={(0.5,2.65)},anchor=north,legend columns=-1,draw=none}, 
                xticklabels={38\%,20\%,0,20\%,40\%,60\%,80\%}, 
                xticklabel style={anchor=north}, 
                tick align=center,
                ytick pos=left,
                bar width = 15pt,
                extra x ticks={0}, 
                extra x tick style={ 
                    grid=major,
                    grid style={black},
                    tick label style={anchor=north},
                },
            ]

            \addplot+[xbar, fill=yellow!90!black, draw=yellow!90!black, mark=none, forget plot] coordinates {
                (#5,0)
                
            };
            
            \addplot+[xbar, fill=red!90, draw=red!90, mark=none, forget plot] coordinates {
               (#6,0)
            };

            \addplot+[xbar, fill=green!70!black, draw=green!70!black, mark=none, forget plot] coordinates {
                (#4,0)
            };

            \addlegendimage{red,fill=red}
            \addlegendentry{Not Useful}
            \addlegendimage{yellow,fill=yellow}
            \addlegendentry{Neutral}
            \addlegendimage{green,fill=green}
            \addlegendentry{Useful}
            \end{axis}
        \end{tikzpicture}
        \label{fig:task-usefulness}
    \end{subfigure}
    \vspace{-0.5cm}
   \caption{(a) Interest for #7 (b) Usefulness Perception of #7}
   \label{#8}
   \vspace{-0.1cm}
\end{figure}
}

\tcbuselibrary{skins,breakable}

\makeatletter
\def\@serieslogo{}
\makeatother

\begin{document}

\title{ChatGPT Inaccuracy Mitigation during Technical
  Report Understanding: Are We There Yet?}

\author{ \IEEEauthorblockN{Salma Begum Tamanna}
    \IEEEauthorblockA{
                    \textit{ salmabegum.tamanna@ucalgary.ca}\\
                    University of Calgary\\
                      Calgary, Canada \\
                     }
    \and
    \IEEEauthorblockN{Gias Uddin}
    \IEEEauthorblockA{\textit{guddin@yorku.ca}\\
                      York University\\
                      Toronto, Canada
                      }
    \and
    \IEEEauthorblockN{Song Wang}
    \IEEEauthorblockA{\textit{wangsong@yorku.ca}\\
                      York University\\
                      Toronto, Canada
                     }
    \and
    \IEEEauthorblockN{Lan Xia}
    \IEEEauthorblockA{\textit{lan\_xia@ca.ibm.com}\\
                      IBM\\
                      Markham, Canada
                     }
    \and
    \IEEEauthorblockN{Longyu Zhang}
    \IEEEauthorblockA{\textit{longyu.zhang@ibm.com}\\
                      IBM\\
                      Markham, Canada
                     }
}

\maketitle
\balance
\begin{abstract}
Hallucinations, the tendency to produce irrelevant/incorrect responses, are prevalent concerns in generative AI-based tools like ChatGPT. Although hallucinations in ChatGPT are studied for textual responses, it is unknown how ChatGPT hallucinates for technical texts that contain both textual and technical terms. We surveyed 47 software engineers and produced a benchmark of 412 Q\&A pairs from the bug reports of two OSS projects.  We find that a
RAG-based ChatGPT (i.e., ChatGPT tuned with the benchmark issue reports) is 36.4\% correct when producing answers to
the questions, due to two reasons \begin{inparaenum}
\item limitations to understand complex technical contents in code snippets like stack traces, and 
\item limitations to integrate contexts denoted in the technical terms and texts.
\end{inparaenum} We present CHIME (\ul{Ch}atGPT \ul{I}naccuracy \ul{M}itigation \ul{E}ngine) whose underlying principle is that if we can preprocess the technical reports better and guide the query validation process in ChatGPT, we can address the observed limitations. CHIME uses context-free grammar (CFG) to parse stack traces in technical reports. CHIME then verifies and fixes ChatGPT responses by applying metamorphic testing and query transformation. In our benchmark, CHIME shows 30.3\% more correction over ChatGPT responses. In a user study, we find that the improved responses with CHIME are considered more useful than those generated from ChatGPT without CHIME.
\end{abstract}

\section{Introduction}\label{sec:introdcution}
The reliability of LLMs is often questioned due to their tendency to produce nonsensical or incorrect outputs, a phenomenon commonly referred to as hallucination \cite{Ji2023, filippova-2020-controlled, Huang2023ASO}. Like any LLM, ChatGPT can also suffer from hallucination issues like inconsistency in responses \cite{jang-etal-2022-becel,Jang2023ConsistencyAO} or factual inaccuracies. These problems can arise even when the model is provided with the context as a document/paragraph.  While progress is made to assess hallucinations in textual data \cite{cohen-etal-2023-lm, Galitsky2023}, we are not aware of how hallucinations can be detected and mitigated for software technical reports that contain both textual and technical terms (e.g., crash dumps, code snippets, etc.) 

This paper studies the detection and mitigation of ChatGPT inaccuracies in technical reports. We pick software bug reports for our study, because bug reports often contain a blend of descriptive text, technical terminology, code references, and snippets of crash/system dumps \cite{OpenJ9Issue18151}.  These documents are crucial for tracking and resolving software issues but can be overwhelming due to their volume and complexity~\cite{InfoTypesOSSDiscussion}. An AI chatbot, trained to understand these reports, may streamline the process by extracting information. But for that, first we need to ensure that the responses from the chatbot are correct.

\textbf{In the first phase of our study}, we conducted a survey of 47 software engineers to understand the types of questions they ask while exploring bug reports and for which they wish for an automated Q\&A tool like a chatbot. We found that developers ask diverse questions during bug exploration, which we could group into five types: \begin{inparaenum}
\item issue analytics,
\item issue trends,
\item issue summary,
\item issue labeling, and
\item issue backlogs.
\end{inparaenum} Based on the survey findings, we produced a benchmark of 412 Q\&A pairs by consulting our industry partner (with whom we conducted regular bi-weekly sessions) and the literature. The Q\&A pairs are collected by analyzing the issue reports of two popular open-source software (OSS). 

\textbf{In the second phase of our study}, we tuned ChatGPT with issue reports from the two studied OSS based on the Retrieval Augmented Generation (RAG) techniques \cite{jiang-etal-2023-active,lewisRag,GuuRag,shuster-etal-2021-retrieval-augmentation}. We then evaluated the correctness of ChatGPT responses against our benchmark. Each question was asked and compared automatically to its expected answer. Correctness was assessed as the ratio of questions whose answers were found as correct. We found that our RAG-based ChatGPT was correct in only 36.4\% cases. For the rest of the questions, it hallucinated by producing incorrect or irrelevant answers. We manually examined each hallucination case and identified two limitations in ChatGPT to process technical documents like bug reports: \begin{inparaenum}
\item limitations to understand complex technical contents in code snippets like stack traces (e.g., when a partial code snippets/crash dump is provided and the question is about determining the cause of the crash by assessing both the crash dump and the textual contents), and 
\item limitations to integrate contexts denoted in the technical terms and texts (e.g., when ChatGPT was required to assess the relationships among multiple metadata and the technical terms).
\end{inparaenum} 

\textbf{In the third phase of our study}, we designed CHIME (\ul{Ch}atGPT \ul{I}naccuracy \ul{M}itigation \ul{E}ngine) to address the above two limitations. The underlying principle in CHIME is that \begin{inparaenum}[(1)]
   \item if we can preprocess the technical reports better and store information relevant to an issue report as a combination of metadata and actual contents and 
   \item then guide the query validation process in ChatGPT with guided iterative prompting approaches, we can address the observed limitations.
\end{inparaenum}  The usage of metadata is found to improve LLM search capabilities \cite{beelen2022metadata}. For us, such metadata could be generated by organizing the mix of textual and technical contents into a structured form. A challenge was how to separate the textual and technical contents and process the code terms within a crash dump and then organize those within a structure. We introduce a novel context-free grammar (CFG) in CHIME to efficiently parse stack traces in technical reports. As for the second principle (i.e., guided prompting for verification), we extended recent similar work on textual content. CHIME verifies and fixes ChatGPT responses by using query transformation \cite{ma2023query} and by extending CoVe \cite{dhuliawala2024chainofverification} with metamorphic testing (MT) \cite{Chen-MTSurvey-CSUR2018}. CoVe is a zero-shot iterative prompting-based query verification technique. We evaluated CoVe's response by further mutating the question using MT, because CoVe may discard correct responses or promote incorrect responses.

We evaluated CHIME using our benchmark. CHIME shows on average 30.3\% improvement over ChatGPT responses by offering more correct answers. In a user study, we find that the improved responses with CHIME are considered more useful than those generated from ChatGPT without CHIME.

Our \textbf{replication package} (\url{https://bit.ly/4fyaMIP}) contains all the data and code developed in the study.

\section{Related Work}\label{sec:related-work}
\subsection{Hallucinations in Large Language Models}

 Extensive studies in the literature identified the causes of hallucinations as sub optimal training, inference \cite{ExposureBias,liu2023exposing,lee-etal-2022-deduplicating, dziri-etal-2021-neural,chang-mccallum-2022-softmax}, and insufficient/low-quality data \cite{Dhingra2019HandlingDR,lin-etal-2022-truthfulqa,Carlini2020ExtractingTD}. Techniques such as bidirectional auto-regressive models \cite{li2023h} and attention-sharpening mechanisms \cite{liu2023h} have been developed to address training-related hallucinations. Inference issues, primarily due to decoding strategies, often result in inaccurate outputs. Strategies like factual-nucleus sampling \cite{Lee2022FactualityEL} and in-context pretraining \cite{shi2023b} are implemented to mitigate these inaccuracies. Challenges posed by flawed data sources introduce biases and inaccuracies into models, stemming from misinformation, duplication biases, and social biases in the training datasets. Mitigating data biases involves manual dataset creation \cite{Radford2019}, integrating high-quality sources such as the Pile \cite{gao2021pile}, and up-sampling factual data \cite{Touvron2023b}. Furthermore, knowledge editing \cite{sinitsin2020editable, yao-etal-2023-editing} and Retrieval-Augmented Generation (RAG) \cite{Jiang2023b, lewisRag, GuuRag, shuster-etal-2021-retrieval-augmentation} are employed to bridge knowledge gaps, utilizing external sources for more accurate text generation.
 
 Our study utilizes RAG-based ChatGPT for technical bug report understanding. We enhance ChatGPT's knowledge base by integrating it with a database of bug reports through RAG methods. We then develop CHIME, which refines both the preprocessing of input data and the validation of RAG-based ChatGPT responses.




\subsection{LLMs for Software Engineering}
In recent years, the application of LLMs has been widely utilized in Software Engineering (SE) tasks, ranging from code analysis to bug detection \cite{SobaniaBugFix,Radford2018ImprovingLU,Lu2021CodeXGLUEAM, Lanchantin2023, Peng2023, Shuster2021, Madaan2023, Malaviya2023, Manakul2023}. Encoder-only models like BERT \cite{Devlin2019BERTPO} and its derivatives, including CodeBERT \cite{feng-etal-2020-codebert} and GraphCodeBERT \cite{Guo2020GraphCodeBERTPC}, excel in processing code. While encoder-decoder models, such as T5 and PLBART excel in understanding semantics of code for tasks like code summarization \cite{ahmad-etal-2021-unified,Raffel2019ExploringTL}. Decoder-only models like the GPT series and specialized versions like CodeGPT and Codex generate direct responses from prompts. However, the specific challenge of mitigating inaccuracies in software technical reports remains unexplored. Addressing this gap, we introduce CHIME to reduce inaccuracies in ChatGPT-generated responses during bug report exploration.
\section{AI Chatbot Needs for Bug Report Exploration}\label{sec:chatbotneed}
To evaluate a chatbot on software technical documents, we needed a benchmark, which at the time of our study was not available. We thus adopted a systematic approach to create such a benchmark. First, we conducted a survey of software developers to produce a catalog of questions that they ask during bug reports. Second, we used the catalog to produce our benchmark (see Section \ref{sec:benchmark}). This section discusses the survey, which answers the following research question (RQ):

\begin{enumerate}[label=\textbf{RQ\arabic{*}.}, leftmargin=30pt]
    \item What types of questions would software practitioners like to ask a chatbot during bug report exploration?
\end{enumerate}

\subsection{Survey Participants} 

\begin{table}[!t]
\centering
\caption{Demography of Survey Participants}
\label{tab:demographics}
\renewcommand{\arraystretch}{1.3} 
\begin{tabular}{l|ccccc |c}
\multicolumn{1}{c}{} & \multicolumn{5}{c}{\textbf{Years of Experience}} \\ \cline{2-7}
\multicolumn{1}{l|}{\textbf{Current Role}} & \textbf{0-3} & \textbf{4-5} & \textbf{6-10} & \textbf{11-15} & \textbf{16-20} & \textbf{Total} \\ \hline
\rowcolor{gray!25}
Developer                                  & 13          & 20           & 4             & 2              & -              & \textbf{39}    \\ 
QA Engineer                                & 1           & 1            & -             & -              & -              & \textbf{2}      \\ 
\rowcolor{gray!25}
Project Manager                            & -           & 1            & -             & -              & -              & \textbf{1}       \\ 
Other                                      & 2           & -            & 2             & -              & 1              & \textbf{5}       \\ \cline{1-7}
\textbf{Total}                             & \textbf{16} & \textbf{22}  & \textbf{6}    & \textbf{2}     & \textbf{1}     & \textbf{47}          
\end{tabular}
\end{table}

\begin{figure}[!t]
    \centering
    \begin{subfigure}{.45\linewidth}
        \centering
        \begin{tikzpicture}
        [pie chart,
        slice type={Daily}{green!30},
        slice type={Weekly}{lime!30},
        slice type={Monthly}{orange!25},
        slice type={Rarely}{red!45},
        pie values/.style={font={\footnotesize}},
        scale=1.1]
        
        \pye{}{70.21/Daily,17.02/Weekly,2.13/Monthly,10.64/Rarely}
        \legend[shift={(-1cm,-1cm)}]{{Daily}/Daily} 
        \legend[shift={(0.5cm,-1cm)}]{{Weekly}/Weekly}
        \legend[shift={(-1cm,-1.3cm)}]{{Monthly}/Monthly}
        \legend[shift={(0.5cm,-1.3cm)}]{{Rarely}/Rarely}
        \end{tikzpicture}
        \caption{}
        \label{fig:freq}
    \end{subfigure}%
    \begin{subfigure}{.45\linewidth}
        \centering
        \begin{tikzpicture}[scale=0.5]
        
            \begin{axis}[
                ybar,
                xtick distance=1.5,
                symbolic x coords={Resolution, Triaging, Reporting, Management },
                xtick=data,
                xticklabel style={font=\large},
                nodes near coords={\pgfmathprintnumber\pgfplotspointmeta\%},
                nodes near coords align={vertical},
                nodes near coords style={font=\large},
                bar width=30pt,
                axis x line*=bottom,
                xtick distance=7,
                y axis line style={opacity=0}, 
                tickwidth=0pt, 
                ytick=\empty,
                width=9cm
                ]
            \addplot[fill=black!30] coordinates {(Resolution,89.36) (Triaging,44.68) (Reporting,36.17) (Management,27.66) };
            
            \end{axis}
        \end{tikzpicture}
        \caption{}
        \label{fig:reason}
    \end{subfigure}
    \caption{(a) Frequency \& (b) Reasons of Bug Report Exploration}
    \label{fig:test}
\end{figure}
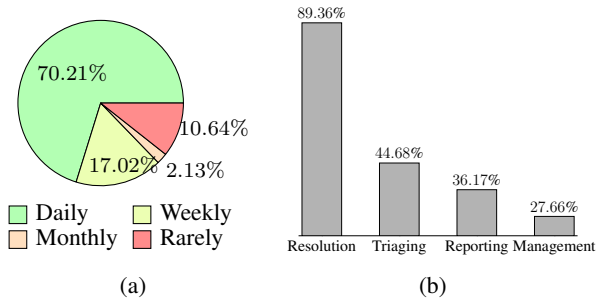

We employed the snowball sampling approach \cite{goodman1961snowball} to recruit participants, resulting in 47 software practitioners. The majority of respondents (83\%) held developer roles. The largest proportion of participants (55\%) reported having 4-5 years of experience in the field. Table \ref{tab:demographics} illustrates the distribution of participants across their roles in the software industry and years of experience. We ensured that all selected participants regularly use Issue Tracking Systems (ITS), such as Jira, GitHub, or in-house systems for several key purposes. Around 89.36\% of respondents used ITS for issue resolution—addressing and for solving bugs; followed by issue triaging which involves prioritizing and assigning issues. Project Management and documenting issues for records or stakeholder communication were also noted. A vast majority of respondents (70.21\%) reported engaging with issue reports daily. Figure \ref{fig:freq} presents the distribution of issue report interaction frequency and Figure \ref{fig:reason} illustrates the survey responses for the primary reason for issue report exploration.

\subsection{Survey Questions}
Before the main survey, we consulted two software professionals from a reputed software company to get insights on the potential tasks that an AI chatbot can support during bug report understanding. Each had 16 and 35 years of experience respectively. We finally settled on five distinct tasks as shown in Table \ref{tab:issue_tasks}). For our survey, we formulated questions to address these tasks only.

\begin{table}[!t]
\caption{Identified Task Types for Bug Report Exploration}
\label{tab:issue_tasks}
\centering
\begin{tabular}{l l p{5cm}}
\toprule
\textbf{T\#} & \textbf{Task Title} & \textbf{Task Description}\\
\midrule
T1 & Issue Analytics & Explores complex details within/across bug reports, including technical jargon, error codes, and contextual nuances, etc.  \\

T2 & Issue Trends                                  & Patterns and trends in bug occurrences.   \\

T3 & Issue Summaries                                    & Summaries of key topics discussed in an issue or across multiple issues.\\

T4 & Issue Labeling   & Inquires labels for bug reports to help organize and categorize them effectively. \\
T5 & Issue Backlog   & Explores whether the issue remains open for long and why\\
\bottomrule
\end{tabular}
\end{table}

During the survey, participants were prompted to envision an AI chatbot with comprehensive access to bug reports. Subsequently, they were asked whether they would like to utilize the chatbot for the specified tasks (T\#) and to rate the perceived usefulness of each task for their work. We condensed the key survey questions into Table \ref{tab:survey_questions} for clarity. All survey questions can be found in our online appendix.

\subsection{Understanding Preferences of Software Practitioners on Identified Task Types (RQ1)}
In Table \ref{tab:survey_questions}, with Key Question 1 (KQ1), we tried to identify the desired chatbot roles during bug report explorations. With KQ2-4, we examined how participants perceived the usefulness of each task type. The questions are summarized in Table \ref{tab:bench-utility}. For each question, we show quantitative evidence from the survey about what the participants thought of the usefulness of the question.
\begin{table}[!t]
\caption{Key Questions (KQ) of the Survey. C/O=Close/Open-ended question}
\label{tab:survey_questions}
\centering
\begin{tabular}{c p{6cm} c}
\toprule
\textbf{KQ\#} & \textbf{Questions} & \textbf{O/C}\\
\midrule
1 & How would you like to utilize a chatbot during bug report exploration?      & O \\

2 & Would you want to use the chatbot for T\#?                                  & C \\

3 & Rate the usefulness of T\# for your work                                    & C \\

4 & Rate the potential usefulness of the following benchmark questions of T\#   & C  \\

\bottomrule
\end{tabular}
\end{table}
\subsubsection{KQ1. Desired Chatbot Roles}
We asked KQ1 as an open-ended question to check whether the participants considered the same five task types that we identified in Table \ref{tab:issue_tasks}. As such, we did not show them KQ2-4 until they answered KQ1.  

Participants desired a tool whose capabilities closely aligned with our predefined tasks (T1–T5). They emphasized the chatbot's potential to analyze issues (T1), such as extracting pivotal details like exceptions or log entries from similar bug reports. According to R43: \textit{ ``Queries for potential duplicate issues could be really helpful''.} The capability to analyze trends (T2) within reported issues to uncover recurring problems was seen as crucial. Participants also noted the importance of the chatbot's ability to summarize (T3) and categorize (T4) issues. The survey responses indicated a significant interest in features that would allow users to query backlogged items (T5), as R43 stated: \textit{ ``Queries for issues with very little recent activity and no clear resolution would be super helpful''.} Apart from these tasks, participants also expressed desires for additional capabilities such as sorting issues based on priority, severity, or difficulty and forecasting resolution times using historical data. We leave support for those tasks as our future work.

\begin{table*}[!t]
\centering
\caption{Perceived Usefulness of Benchmark Questions Presented in the Survey.\\
\color{red}\rule{4pt}{6pt} \color{black} Not Useful \color{yellow}\rule{4pt}{6pt} \color{black} Neutral \color{green}\rule{4pt}{6pt}\color{black}  Useful}

\begin{tabular}{ l p{10cm} l}
\toprule
\textbf{Q\#} & \textbf{Question}                                                        & \textbf{Perceived Usefulness of T\# (KQ4)} \\ 
\midrule

\multicolumn{3}{l}{\cellcolor{gray!25}{\textbf{T1 - Issue Analytics} | Extracting Information from Issue Details or Find Similarities Among Issues}} \\
Q1.1 & Is there a stack trace provided in issue 123, and can you summarize it?          & \customBarChart{-8.51}{-10.64}{78.72}\\ 
Q1.2 & Where in the code does the exception in issue 123 occur?                         & \customBarChart{-10.64}{-10.64}{76.60}\\
Q1.3 & What is the exception reported in issue 123?                                     & \customBarChart{-12.77}{-12.77}{74.47}\\ 
Q1.4 & How many tests failed as reported in issue 123?                                  & \customBarChart{-4.26}{-27.66}{65.96}\\
Q1.5 & Which environment is associated with the exception reported in issue 123?        & \customBarChart{-36.17}{-6.38}{57.45}\\ 
\hline

Q1.6 & Are there any issues similar to issue 123?                                       & \customBarChart{-10.64}{- 4.26}{82.98}\\ 
Q1.7 & Find duplicate reports of the X error (or other) in 'A' module                   & \customBarChart{-14.89}{-2.13}{80.85}\\
Q1.8 & Find all similar issues related to X failures]                                   & \customBarChart{-14.89}{-2.13}{80.85}\\
Q1.9 & Identify any performance degradation issues reported on last month               & \customBarChart{-25.53}{-4.26}{70.21}\\
Q1.10& Has there been a report of a crash on a X machine running the "A" Test recently? & \customBarChart{-25.53}{- 6.38}{68.09}\\ 
\midrule

\multicolumn{3}{l}{\cellcolor{gray!25}{\textbf{T2 - Issue Trend} | Detect and Analyze the Trends and Patterns among Issues}} \\
Q2.1 & What are the frequently encountered errors in the nightly builds?                & \customBarChart{-14.89}{-8.51}{76.60}\\ 
Q2.2 & What are the recurring themes in bug reports post the latest OS update?          & \customBarChart{-19.15}{-6.38}{74.47}\\      
\midrule

\multicolumn{3}{l}{\cellcolor{gray!25}{\textbf{T3 - Issue Summary} | Obtain a Comprehensive Overview of Reports Selected by Different Criterion}} \\
Q3.1 & List all issues related to an X feature and their current status                  & \customBarChart{-21.28}{- 2.13}{76.60}\\ 
Q3.2 & Generate a report detailing the distribution of issues across different project modules & \customBarChart{-19.15}{- 4.26}{76.60}\\ 
Q3.3 & Compile a summary of unresolved issues not older than 60 days                    & \customBarChart{-21.28}{- 4.26}{74.47}\\
Q3.4 & Can you generate a summary of all issues tagged as 'bug' in the last 30 days?    & \customBarChart{-19.15}{-8.51}{72.34}\\ 
Q3.5 & Create a summary of user-reported issues versus internally identified issues     & \customBarChart{-36.17}{- 10.64}{51.06}\\      
\midrule

\multicolumn{3}{l}{\cellcolor{gray!25}{\textbf{T4 - Issue Label} | Provide Suggestions for Categorizing and Tagging Issues with Appropriate Labels}} \\
Q4.1 & Suggest existing labels to tag issue 123                                         & \customBarChart{-27.66}{- 4.26}{68.09}\\ 
Q4.2 & Can you recommend labels for performance-related issues?                         & \customBarChart{-23.40}{-8.51}{68.09}\\  
\midrule

\multicolumn{3}{l}{\cellcolor{gray!25}{\textbf{T5 - Issue Backlog} |  Analyze Unresolved Issues Reported but not yet Addressed }} \\
Q5.1 & Are there any long-standing issues that have been consistently postponed?        & \customBarChart{-17.02}{-8.51}{70.21}\\
Q5.2 & Find issues that have not been assigned to any milestone but are older than 60 days & \customBarChart{-17.02}{-10.64}{68.09}\\
Q5.3 & List issues that have missed two or more release cycles                          & \customBarChart{-17.02}{-14.89}{63.83}\\
Q5.4 & Identify issues with no activity in the last 30 days.                            & \customBarChart{-23.40}{-10.64}{61.70}\\
\bottomrule
\end{tabular}

\label{tab:bench-utility}
\end{table*}

\customTwoBarChart{72.34} {-17.02} {-10.64}{72.34}{-23.4}{-4.26}{T1}{fig:q2q3T1}
\subsubsection{KQ2-4. Issue Analytics (T1)} 
 The survey results reveal a strong preference for the chatbot’s analytical capabilities, particularly in the context of analyzing multiple issues, with more than 80\% of participants expressing interest in utilizing the chatbot for detecting similar or duplicate issues and finding it useful; while 72.34\% participants value the chatbot's utility in analyzing individual issues as illustrated in Figure \ref{fig:q2q3T1}.

In Table \ref{tab:bench-utility}, we show 10 questions under T1 that each participant assessed. On average, 73.62\% of participants marked these useful. The capability to identify and summarize stack traces (Q1.1) within the single issue analysis domain was highly valued, evidenced by a utility score of 78.72\% and remark from respondent R11, \textit{``Summary of stack trace is a good idea``}.  In comparison, the importance of determining the environment linked to an exception (Q1.2) was rated lower, at 57.45\%. 
For the analysis of multiple issues, the ability to find similar issues (Q1.3) was highly valued at 82.98\%, as quoted by R09, \textit{'They are all extremely time-consuming when done manually. A chatbot will definitely help with this.'}


\customTwoBarChartwithlimit{87.23} {-12.77} {0.00} {85.11} {-10.64} {-4.26} {T2} {fig:q2q3T2}
\subsubsection{KQ2-4. Issue Trend (T2)}  
87.2\% of participants expressed interest in using this feature and 85.1\% found the corresponding questions on identifying and analyzing trends within bug reports useful (see Figure \ref{fig:q2q3T2}). In Table \ref{tab:bench-utility}, we show two questions that we asked under this task. Both received a favorable response. Respondent R31 noted, \textit{``By focusing on recurring errors and themes, these questions provide valuable insights that can guide decision-making, resource allocation, and issue resolution efforts.``} When participants were asked about the utility of chatbots in identifying frequently encountered errors in the nightly builds of their development environment (Q2.1), e.g., in CI/CD pipelines, 76.60\% perceived this functionality as useful. 


\customTwoBarChartwithlimitanother{80.85} {-17.02} {-2.13} {70.21} {-25.53} {-4.26} {T3} {fig:q2q3T3}\subsubsection{KQ2-4. Issue Summary (T3)}
80.6\% of participants were keen on a chatbot to produce summaries of issues and 70.2\% considered the asked questions useful for efficiently understanding and resolving software issues. In Table \ref{tab:bench-utility}, regarding the chatbot's ability to report on how issues are distributed across different project modules (Q3.1), 76.6\% found this function useful. On the other hand, the feature for distinguishing between user-reported and internally identified issues (Q3.2) was seen as useful by 51.06\% of participants, indicating a notable but more moderate interest in differentiating the sources of issues.\\


\customTwoBarChart{70.21} {-23.40} {-6.38} {61.70} {-23.40} {-14.89} {T4} {fig:q2q3T4}\subsubsection{KQ2-4. Issue Labeling (T4)} 
70.2\% of participants are interested in leveraging chatbots for the task of issue labeling (see Figure \ref{fig:q2q3T4}). However, it's worth noting that this task received the lowest percentage of perceived usefulness (61.70\%) compared to others. Regarding the chatbot's ability to suggest appropriate labels for an issue (Q4.1) and to recommend labels for performance-related issues (Q4.2), about 68\% of respondents considered these features to be useful.


\customTwoBarChart{63.83} {-19.15} {-17.02} {65.96} {-14.89} {-19.15} {T5} {fig:q2q3T5}\subsubsection{KQ2-4.  Issue Backlog (T5)}
The management of Issue Backlogs is an essential aspect of software development. R16 highlighted the challenge: \textit{``Sometimes change of priorities pushes issue out of find and stay unresolved for days. So it is good to find out long-running or inactive issues.''} Despite its importance, this task garnered the least interest (63.8\%) among all tasks for potential chatbot utilization. Notably, T5 records the highest percentage of "No" responses (17.02\%) regarding interest and "Not Useful" perceptions (19.15\%). 70.2\% of respondents see value in identifying long-standing, postponed issues (Q5.1) but interest slightly drops to 61.70\% for detecting issues with no recent activity over the last 30 days (Q5.2). 


\mybox{\textbf{Summary of RQ1.} 
When examining software practitioners' preferences for AI chatbot capabilities in exploring bug reports, the identification of similar issues and the analysis of recurring error trends were highly favored. In contrast, capabilities related to categorizing issues and handling pending bugs were deemed less critical. 
}

\section{A Benchmark of Q\&A Pairs to Evaluate AI Chatbots for Bug Report Exploration}\label{sec:benchmark}
 In Table \ref{tab:bench-utility}, we showed a catalog of 23 questions that we validated with our survey participants and for which they wished for chatbot support. Each question is a template, which can be used to produce multiple similar questions.Based on the question templates in Table \ref{tab:bench-utility}, we produced a total of 412 questions from the issue reports of two popular OSS repos, OpenJ9 and ElasticSearch. OpenJ9 was chosen due to its alignment with our industrial partner. ElasticSearch \cite{es1,es2,es3} is frequently referenced in academic studies. We then produced an answer to each question by assessing the two OSS repos and by consulting among the authors. Four authors (the first two and last two) engaged in many hours of discussions that spanned over six months (both in-person and over formal presentations). The last two authors are also among the maintainers of OpenJ9. Given the benchmark was created via mutual discussion, we did not compute any standard agreement analysis metrics. 

We created the benchmark by selecting 80 complex issues (40 from each repository). Following Deeksha et al. \cite{InfoTypesOSSDiscussion}, we define an issue as complex if it is excessively long and/or it has stack traces. We picked issues within the last year of our analysis because those are likely to be explored more by developers. We sorted issues by length and selected 40 issues with stack traces (per repo). Following the standard chatbot evaluation process, we contained three types of answers: binary \cite{Rasool2023EvaluatingLO}, factual \cite{Wang2023SurveyOF,Polak2023ExtractingAM}, and summary \cite{Basyal2023TextSummarization}.

The binary (i.e., Query Type = Y/N) queries have questions with answers as Y/N. These are designed to assess the chatbot's accuracy in identifying clear-cut, definitive binary decisions based on information available in bug reports; such as the presence of a particular error code or the applicability of a specific scenario. To verify a chatbot response for these queries, we simply need to check for Y/N in their responses and match those against the benchmark answer.

 \qabox{Q1. Type: Y/N. Source: ElasticSearch}{\textbf{Question:} Is there any issue similar to issue 100071?}{\textbf{Expected Answer:} No.}

 The factual (i.e., Query Type = Factual) queries assessed the chatbot's ability to extract concrete information from bug reports, such as identifying, retrieving, and presenting specific details from the dataset, such as error messages, stack traces, configuration settings, etc. Like binary queries, this method also allows for a straightforward assessment of the chatbot's accuracy, and thus direct matching can be used for verification.

    \qabox{Q2. Type: Factual. Source: ElasticSearch}{\textbf{Question:} What existing label is recommended for issues that need immediate triaging?}{ \textbf{Expected Answer:} 'needs:triage'}

The summary-based (i.e., Query Type = Summary) queries challenge the chatbot to engage in deeper analysis and synthesis of data.
These queries require the chatbot to identify patterns and even to propose potential solutions based on the analysis of multiple data points. Since these queries demand a synthesis of information and provide insights or summaries, we need a similarity analysis between a response and the expected answer for verification.

    \qabox{Q3. Type: Summary. Source: ElasticSearch}{\textbf{Question:} Summarize similarities between issues 103072
\& 103344}{\textbf{Expected Answer:} Issues 103072 and 103344 both involve test failures within the LearningToRankRescorerIT class. The root cause of these failures stems from a named\_object\_not\_found\_exception and x\_content\_parse\_exception, resulting in ...}

\begin{table}[!h]
\centering
\caption{Distribution of Benchmark Questions over Survey-identified Tasks from Table \ref{tab:bench-utility}.}

\begin{tabular}{l l}
\toprule
\textbf{T\#} & \color{gray!20}\rule{4pt}{5pt} \color{black}  Y/N \color{gray!40}\rule{4pt}{5pt} \color{black} Fact \color{gray!60}\rule{4pt}{5pt}\color{black}  Summarization \\ 
\midrule
T1 - Issue Anlys(S) & \custombarchartwithvalue{48}{140}{24}{0.024} \\
T1 - Issue Anlys(M) & \custombarchartwithvalue{12}{20}{8}{0.1}\\ 
T2 - Issue Trend & \custombarchartwithvalue{16}{16}{8}{0.1}\\ 
T3 - Issue Summary &  \custombarchartwithvalue{0}{8}{32}{0.1}\\
T4 - Issue Labeling & \custombarchartwithvalue{12}{20}{8}{0.1}\\
T5 - Issue Backlog & \custombarchartwithvalue{12}{24}{4}{0.1}\\
\bottomrule
\end{tabular}
 
\label{tab:queryDistribution}
\end{table}

Table \ref{tab:queryDistribution}  shows the distribution of question types—Yes/No, Fact, and Summarization—across OpenJ9 and ElasticSearch, totaling 206 questions per project. OpenJ9 and ElasticSearch have a similar overall structure, with a strong emphasis on factual questions (114 for OpenJ9, 114 for ElasticSearch), followed by binary (yes/no) and summarization questions. Our online appendix contains details about each of the 412 questions and how each question maps to our catalog of 23 survey questions.

\section{Effectiveness of ChatGPT on the Benchmark} \label{sec:rag}
In this section, we answer the following research question: 

\begin{enumerate}[label=\textbf{RQ\arabic{*}.}, leftmargin=30pt,start=2]
    \item How effective is a RAG-enhanced ChatGPT to answer to the benchmark questions while exploring the corresponding bug reports?
\end{enumerate}

\subsection{Approach}
\begin{figure}[h]
\centering
\includegraphics[scale=0.38]{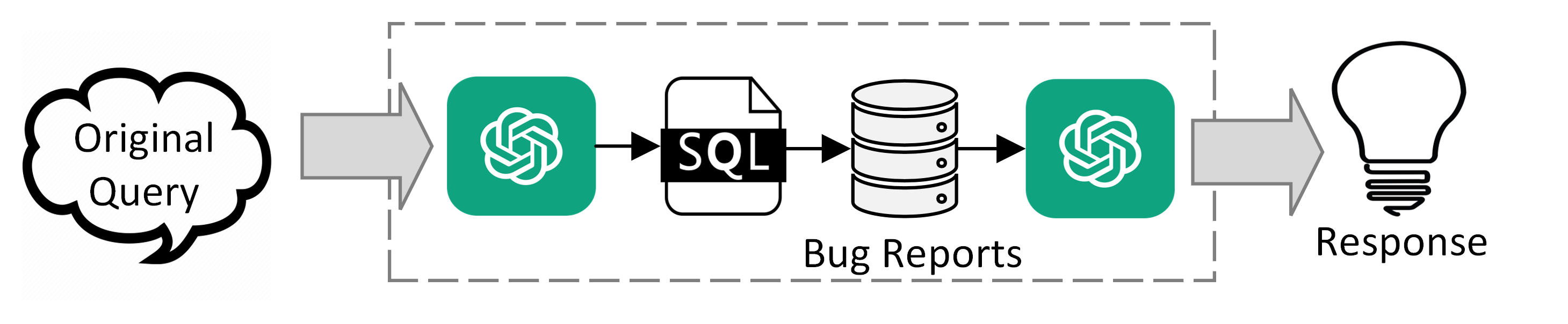}
\caption{Pipeline of the RAG-based ChatGPT}
\label{fig:rag}

\end{figure}

The RAG architecture combines ChatGPT with an external knowledge retriever to provide responses to queries. This framework utilizes external database sources, primarily issue reports with structural data and metadata fetched by the GitHub API. Figure \ref{fig:rag} illustrates the pipeline for this. 
It functions by first retrieving pertinent information from the database based on the input query. This step is crucial as it aligns the model's focus with the most relevant data. Then, the augmented data from the retrieval step are combined with the inherent generative capabilities of ChatGPT to help ChatGPT provide high-quality responses. We used ChatGPT 3.5-turbo within LangChain framework  \cite{langchain} to implement this pipeline. We used a temperature setting of 0. A temperature value above 0 produces slightly different answers to a prompt across multiple runs, which is unnecessary when we expect consistent answers from ChatGPT. We ran it multiple times on our benchmark dataset to ensure that the answers were indeed consistent across multiple runs.

We measure the effectiveness of the above RAG-based ChatGPT on our benchmark by calculating correctness (C): 
\begin{equation}\label{eq:correctness}
    C = \frac{\text{Number of Correct Responses}}{\text{Total Number of Queries}} \times 100\%
\end{equation}
Correctness analysis involved comparing the generated responses against the predefined correct answer for a query in the benchmark. We adopted the following automated approaches to measure the correctness of the responses for the three types of queries in our benchmark (i.e., Y/N, Fact, and Summaries).

For evaluating yes/no responses, we used a zero-shot classification approach, enabling us to automatically determine if detailed answers from the chatbot align with a simple ``Yes'' or ``No'' expectation. For evaluating factual query responses, we combine two approaches: direct comparison of extracted information (such as issue numbers) and semantic similarity assessment for non-listed facts. First, we extract and compare key factual elements. If the response and expectation directly match or share common elements, the correctness is assessed accordingly. For responses without explicit factual elements, we utilize the SentenceTransformer library, employing the \textit{all-MiniLM-L6-v2} model \cite{huggingface} to encode the actual and expected answers into embeddings. Subsequently, we gauge the semantic similarity between these embeddings by computing the cosine similarity \cite{Rahutomo2012SemanticCS} and assessing how closely the actual answer aligns with the expected fact, considering nuances in phrasing and context. For evaluating Summarization queries, we compute the semantic similarity like before between the actual summary provided by ChatGPT and the expected summary. Based on empirical observations (see Section \ref{sec:threats}), we used a similarity threshold of 0.7.

\subsection{Results}\label{sec:rag-results}
The RAG-based pipeline achieved 36.4\% accuracy in our benchmark (see Table \ref{tab:ragresult}). We manually assessed each of the 262 incorrect answers to determine the causes of its incorrectness. Given that ChatGPT is a black-box model, our assessment is based on the nature of the questions asked and the provided answers. We observed two limitations in ChatGPT: 
\begin{enumerate}[label=\textbf{L\arabic{*}.}]
    \item Limitations in Understanding Complex Technical Content (42.7\% cases).
    \item Contextual Understanding Challenges (57.3\% cases).
\end{enumerate}

\begin{table}[!h]
    \centering
    \caption{Correctness of RAG Based ChatGPT}
    \label{tab:ragresult}
    \begin{tabular}{lcccc}
        \toprule
        \textbf{T\#} & \textbf{Y/N} & \textbf{Fact} & \textbf{Sum} & \textbf{Total} \\
        \midrule \rowcolor{gray!25}
        Total & 49.0\% & 30.7\% & 36.9\% & 36.4\% \\
        \midrule
        T1 - Issue Anlys(S) & 41.7\% & 32.1\% & 33.3\% & 34.4\% \\
        T1 - Issue Anlys(M) & 50.0\% & 15.0\% & 0.0\% & 22.5\% \\
        T2 - Issue Trend & 50.0\% & 12.5\% & 37.5\% & 32.5\% \\
        T3 - Issue Summary & 0.0\% & 75.0\% & 43.8\% & 50.0\% \\
        T4 - Issue Labeling & 58.3\% & 40.0\% & 62.5\% & 50.0\% \\
        T5 - Issue Backlog & 66.7\% & 25.0\% & 25.0\% & 37.5\% \\
        \bottomrule
    \end{tabular}
\end{table}

The challenge in understanding technical contents (L1) occurred when ChatGPT faced challenges in deeply understanding and accurately processing highly technical content, particularly when it involved intricate programming/technical concepts. For example, when asked to provide the root location of the exception encountered in issue 18151; it gives a non-useful answer; even though the issue provides a stack trace.
    \qabox{Q4. Asking ChatGPT - OpenJ9}{\textbf{Question:} Where in code exception of issue 18151 occur?} { \textbf{Incorrect Answer:} The exception of issue 18151 occurs in the body of the issue.}

 Again, for asking to find the line number where the error occurs, ChatGPT just gives an irrelevant answer by showing the filename.
    \qabox{Q5. Asking ChatGPT - OpenJ9}{\textbf{Question:} Which line in CharacterDataLatin1 class triggered ArrayIndexOutOfBoundsException in issue 17063?} { \textbf{Irrelevant Answer:} The line of code that is "java/lang/String/ToLowerCase.java".}

The Contextual Understanding Challenges category (i.e., L2) shows a fundamental difficulty in ChatGPT to process and respond to queries about technical issues. This category primarily involves the model's struggles with:
\begin{itemize}[leftmargin=10pt]
\item Integrating and interpreting the context in which queries are made. Some context is explicitly stated within the query or the referenced issue, such as a specific error message or stack trace. Other times, the context is implicit, requiring the model to infer based on its broader knowledge or related data points. Handling ambiguous or insufficiently detailed queries necessitates the chatbot to fill in the gaps with assumptions or inferred knowledge. Not all contextual information holds equal relevance to a given query, requiring the chatbot to prioritize the most pertinent context based on the nuances of the query.   

\item Adapting to the technical conventions of specific domains. Technical domains often have their own conventions for documentation, communication, and issue tracking. For instance, understanding that a particular label in an issue tracking system denotes the responsible team, requires domain-specific knowledge that the AI must possess. On asking to find a responsible team for an issue, ChatGPT searches on the assignee list, but the team details are on the issue labels. Due to this lack of contextual information, it fails to answer the question.  

\qabox{Q6. Asking ChatGPT - ElasticSearch}{\textbf{Question:} Which team is responsible for issue 104160?} { \textbf{Incorrect Answer:} The team responsible for issue 104160 is not specified in the database.}

\item Even when the relevant context is identified, retrieving and applying it accurately to generate a response is challenging. This includes understanding the specific ways in which information is structured or presented within data sources and how it relates to the user's query.  Effectively bridging this gap is crucial for generating accurate and contextually appropriate responses. 
\end{itemize}

\mybox{\textbf{Summary of RQ2.} 
A RAG-based ChatGPT showed an average correctness of 36.4\% on our benchmark. The pipeline encountered challenges in comprehending complex technical content and grasping contextual nuances, leading to inaccuracies in its responses.
}

\section{CHIME: {Ch}atGPT {I}naccuracy {M}itigation {E}ngine} \label{sec:chime}
Our observations in Section \ref{sec:rag} of ChatGPT limitations contributed to the design of CHIME, as a suite of techniques to detect and fix incorrectness in ChatGPT responses. The underlying principle of CHIME is that by offering ChatGPT with a more structured representation of bug reports and by applying a systematic approach to assess ChatGPT responses, we can address the two limitations we observed in Section \ref{sec:rag-results}.
A more structured representation of bug reports can be achieved if we can process the different technical and textual terms properly and store those in a structured way, e.g., in a database with metadata offering more information about those terms. A systematic approach to verify the responses can be achieved by applying/adapting the techniques of guided iterative prompting of LLM responses that are used in the literature for textual content. As such, we designed  to preprocess the inputs (both the bug report and the query) and to verify the ChatGPT responses. 

CHIME treats ChatGPT as an API, where the inputs (questions) and outputs (answers) are processed for inaccuracy detection and mitigation. We can use another LLM as an API in CHIME and apply all the techniques we developed. Doing so would simply require changing the API endpoints to point to the other LLM within the LangChain toolkit. CHIME will need to be updated significantly while using multi-modal LLMs, e.g., to process/validate modalities other than texts, etc.

\begin{figure}[!h]
\centering
\includegraphics[scale=0.4]{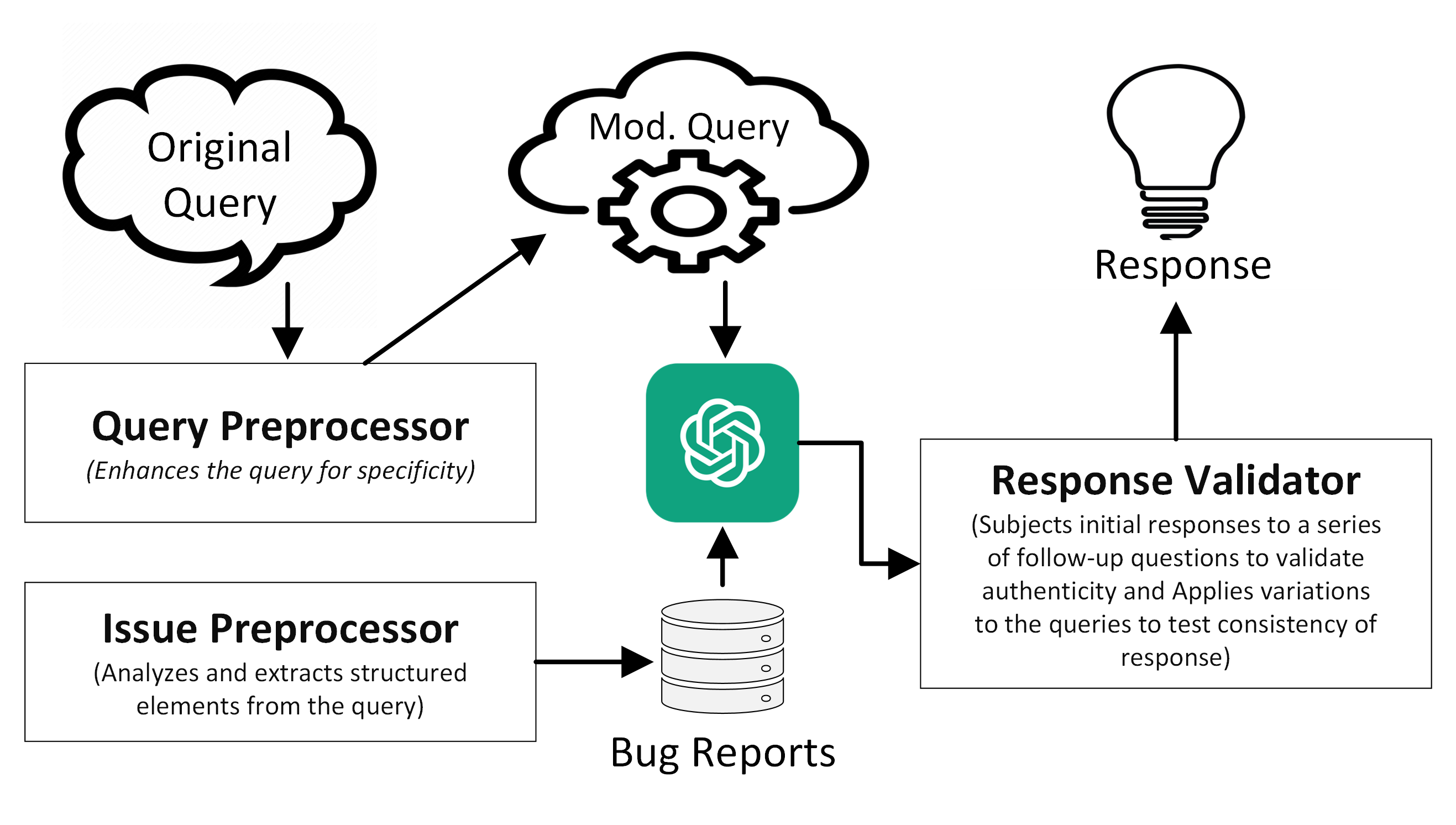}
\caption{The Architecture of CHIME}
\label{fig:CHIME}

\end{figure}

In Figure \ref{fig:CHIME}, we show the architectural diagram of CHIME. We apply an `Issue Preprocessor' component to parse crash dumps and other technical terms. Issue Preprocessor employs Context-Free Grammars (CFGs) to interpret complex technical data, addressing ChatGPT's {limitations in understanding complex technical content}. To address the second limitation (i.e., Contextual understanding challenges), we introduce two more components in CHIME. 
First, we apply `Query Preprocessor' component to decipher users' intents. Second, we designed the `Response Validator' component, which evaluates the accuracy in a response using a combination of two techniques: Chain of Verification (CoVe) \cite{dhuliawala2024chainofverification} and Metamorphic Testing (MT) \cite{Chen-MTSurvey-CSUR2018}. A query goes through each of these components. We discuss these components below.

\subsection{Issue Preprocessor }  We developed a CFG as shown in Listing \ref{lst:stack_trace_grammar} which supports parsing stack traces to the level of individual code elements. Stack traces can also contain auxiliary information like timestamps and memory addresses (e.g., crash dumps). We preprocessed these auxiliary information and applied our CFG to parse the code elements in a Java stack trace. Finally, our issue preprocessor stored the code blocks in the database by separating the code blocks from the textual contents. The issue preprocessor saves the following information by parsing an issue report: title, body, stack trace, and other metadata (creation date, status, etc). When ChatGPT, enhanced with Issue Preprocessor, is presented with the query regarding the type of exception in issue 18151 (Q4) or 17063 (Q5), it correctly identifies the location of the exception detailed in the stack trace.

\begin{lstlisting}[language={},caption={CFG Grammar for Stack Traces},breaklines=true, label=lst:stack_trace_grammar, 
  prebreak={},postbreak={},basicstyle=\small]
Root ::= StackTraceElems
StackTraceElems ::= StackTraceElem StackTraceElems | StackTraceElem
StackTraceElem ::= ExceptionElems | CodeDetails
ExceptionElems ::= ExceptionElem ExceptionElems | ExceptionElem
ExceptionElem ::= ExceptionType | ExceptionMessage 
CodeDetails ::= ClassElem | MethodElem | FileElem | LineElem
\end{lstlisting}

\qabox{Asking CHIME with Q4 from Section \ref{sec:rag}. OpenJ9}{\textbf{Question:} Where in the code exception of issue 18151?} { \textbf{Correct Answer:} The exception in issue 18151 can be triggered at line 98 in the JvmErgonomics.java file. }

\qabox{Asking CHIME with Q5 from Section \ref{sec:rag}. OpenJ9}{\textbf{Question:} Which code in CharacterDataLatin1 triggered the ArrayIndexOutOfBoundsException in issue 17063?} { \textbf{Correct Answer:} Line 72 in the CharacterDataLatin1 class.}


\subsection{Query Preprocessor}  A user query may not provide enough details, leading ChatGPT to generate responses based on faulty assumptions or context. We leveraged the query transformation algorithm from \cite{ma2023query} that strategically rephrases and modifies a query. Each query goes through this component for error correction as follows. We prompt ChatGPT with instructions on how to transform a query based on the question type. For example, for a Yes/No query type, we instruct ChatGPT as follows ``Change `is/are/have there issues' to check if there are any issues with the provided condition''. For a summarization type question, the instruction is ``Summarize the contents from issue title, exceptions, body, and labels''. 
We then provide some example transformations (in a few-shot setting) to clarify the instruction, like the following.

\qabox{Query Preprocessor.}{\textbf{Original Query:} Are issue 18102 and 18669 similar?} { \textbf{Transformed Query:} Compare the exceptions, stack traces, and descriptions of issues 18102 and 18669 to determine similarity.}

Finally, GPT answers based on the transformed query. For example, when presented with the query, Q6 from Section \ref{sec:rag}, the Query Preprocessor directs the question with additional instructions: "Summarize the team responsible for issue 104160 based on the assignee and labels." This instruction results in generating a more precise SQL query to our database (we used LangChain SQL engine + RAG ChatGPT pipeline).
\qabox{Asking CHIME with Q6 from Section \ref{sec:rag}. ElasticSearch}{\textbf{Question:} Which team is responsible for issue 104160?} { \textbf{Correct Answer:} The team responsible for issue 104160 is the ML team, with no specific assignee and labels indicating it is related to ML.}


\subsection{Response Validator} 

The third component of CHIME is added to validate initial responses from RAG-based ChatGPT. We employed two strategies sequentially, CoVe \cite{dhuliawala2024chainofverification} and Metamorphic Testing (MT) \cite{Chen-MTSurvey-CSUR2018}.  Drawing inspiration from investigative interrogation techniques, CoVe's methodology is designed to test the consistency and authenticity of ChatGPT's responses through a series of challenge questions. CoVe's process can be unfolded into three distinct stages \begin{inparaenum} \item Upon receipt of a query, ChatGPT formulates an initial response based on its interpretation of the query within the context of available data. \item CoVe then autonomously generates a set of follow-up questions related to the initial response. CoVe dynamically determines the number of follow-up questions. These questions aim to clarify its preceding statements. \item The subsequent responses are analyzed for consistency with the initial response. Any observed contradictions signal the need for a review and adjustment of the initial response, leading to the refinement or generation of a new, more accurate answer. \end{inparaenum} 

\begin{lstlisting}[language=TeX, basicstyle=\small, breaklines=true, numbers=none,caption={How CoVe validates in CHIME}, prebreak={},postbreak={},label={lst:coveexample}, escapeinside={(*@}{@*)}]
(*@\textbf{Initial Question:}@*) Is issue 18102 and 18669 similar?
(*@\textbf{Initial ChatGPT Response:}@*) No, issue 18102 and 18669 are not similar.

(*@\textbf{Follow-up Questions Generated by CoVe}@*)
1. Can issue 18102 and 18669 be considered similar based on their descriptions and reported problems?
2. Have there been any interactions or dependencies between the teams working on issue 18102 and 18669 that could indicate similarity?
3. Are the issues 18102 and 18669 related to the same error?

(*@\textbf{ChatGPT Responses to Follow-up Questions}@*)
1. Yes, based on the descriptions and reported problems, they can be considered similar as both involve ArrayIndexOutOfBoundsException.
2. Yes, there have been interactions between the teams working on issue 18102 and 18669.
3. No, issues 18102 and 18669 are not related to same error.

(*@\textbf{Final Response:}@*) Yes, issue 18102 and 18669 are similar. 
\end{lstlisting}
However, CoVe can discard correct responses (if not found in responses to challenge questions) or promote incorrect responses (if responses are found both in base and challenge questions). We thus refine CoVe with MT as follows.

Using CoVe to verify ChatGPT's initial responses, we get an intermediate response which is then passed for consistency verification to MT. Central to MT is the concept of metamorphic relations (MRs), which involve applying transformations to input queries to anticipate changes in output. In CHIME, this approach is leveraged to introduce nuanced variations to the queries for testing the consistency of its responses under diverse conditions. A specific implementation of MRs in CHIME involves Sentence-Level Mutation, where equivalent questions are generated to test. This process includes \begin{inparaenum} \item using predefined MRs to subtly alter the phrasing or focus of the original queries. In our implementation, MT generates three mutated questions each time in CHIME. For instance, the original query might be ``What causes error X in module Y?'', and its mutated equivalent could be ``Why does error X occur in module Y?''. \item The responses ChatGPT provides for both the original and mutated queries are compared to assess consistency with MR expectations. If an MR implies that adding specific details should refine the response, the evaluation checks for this level of refinement. After the refinement via MR, CHIME generates the final response for the users. \end{inparaenum} The user in CHIME does not see the mutated questions from CoVe or the mutated questions from MRs. Instead, the user only sees the final response from CHIME. However, CHIME can print the mutated questions to the user if needed.

\begin{lstlisting}[language=TeX, basicstyle=\small, breaklines=true, numbers=none,caption={How CoVe + MT validates in CHIME}, prebreak={},postbreak={} ,label={lst:rvexample},escapeinside={(*@}{@*)}]
(*@\textbf{Initial Question:}@*) Does Elasticsearch require the UseG1GC option to be present during issue 18151 startup stage?
(*@\textbf{Initial ChatGPT Response:}@*) No, Elasticsearch does not require the UseG1GC option to be present during its startup stage in issue 18151.

(*@\textbf{Follow-up Questions Generated by CoVE}:@*)
1. Is Elasticsearch the software mentioned in the response?
2. Is the UseG1GC option not required?
3. Is issue 18151 not requiring the UseG1GC option?

(*@\textbf{ChatGPT Responses to Follow-up Questions}:@*)
1. Yes, Elasticsearch is mentioned in the response.
2. No, the UseG1GC option is not required.
3. No, issue 18151 is not requiring the UseG1GC option.

(*@\textbf{Intermediate Response from CoVE}:@*) No, the UseG1GC option is not required during Elasticsearch's startup stage.

(*@\textbf{Mutated Questions Generated by MT}:@*)
1. Is the UseG1GC option necessary for Elasticsearch to be present during its startup stage in issue 18151?
2. Must the UseG1GC option be included during Elasticsearch's startup stage for issue 18151?
3. Is it required to have the UseG1GC option present during Elasticsearch's startup stage for issue 18151?

(*@\textbf{ChatGPT Responses to Mutated Questions}:@*)
1. No, the UseG1GC option is not necessary for Elasticsearch's to be present during its startup stage in issue 18151.
2. Yes, the UseG1GC option must be included during Elasticsearch's startup stage for issue 18151.
3. Yes, it is required to have the UseG1GC option present during Elasticsearch's startup stage for issue 18151.

(*@\textbf{Final Response:}@*) Yes, it is required to have the UseG1GC option during Elasticsearch's startup stage .
\end{lstlisting}


\section{Effectiveness of CHIME}\label{sec:result}
We evaluate CHIME by answering three RQs:
\begin{enumerate}[label=\textbf{RQ\arabic{*}.}, leftmargin=30pt,start=3]
    \item How well can CHIME fix inaccuracies in ChatGPT responses?
    \item How well do individual components in CHIME perform over ChatGPT?
    \item Would responses from CHIME be favored like those from ChatGPT when both are correct?
\end{enumerate}
RQ3 and RQ4 investigate whether CHIME and its components can fix inaccuracies in ChatGPT while we use our benchmark. Similar to RQ2, we use the correctness metric from Equation \ref{eq:correctness} to answer RQ3 and RQ4. RQ5 assesses the usability of CHIME when it is used by developers instead of a RAG-based ChatGPT. We conduct a user study to answer RQ5.

\begin{table}[!h]
    \centering
    \caption{Correctness of CHIME. Column `Improv' shows percent improvement over RAG-based ChatGPT}
    \label{tab:CHIMEresult}
    \begin{tabular}{lccccc}
        \toprule
        \textbf{T\#} & \textbf{Y/N} & \textbf{Fact} & \textbf{Sum} & \textbf{Total} & \textbf{Improv} \\
        \midrule \rowcolor{gray!25}
        Total & 80.0\% & 61.4\% & 65.5\% & 66.7\% & +30.3\% \\
        \midrule
        T1 - Issue Anlys(S) & 83.3\% & 67.1\% & 66.7\% & 70.8\% & +36.3\% \\
        T1 - Issue Anlys(M) & 58.3\% & 30.0\% & 50.0\% & 42.5\% & +20.0\% \\
        T2 - Issue Trend & 68.8\% & 43.8\% & 50.0\% & 55.0\% & +22.5\% \\
        T3 - Issue Summary & 0.0\% & 87.5\% & 68.8\% & 72.5\% & +22.5\% \\
        T4 - Issue Labeling & 83.3\% & 60.0\% & 87.5\% & 72.5\% & +22.5\% \\
        T5 - Issue Backlog & 100.0\% & 58.3\% & 50.0\% & 70.0\% & +32.5\% \\
        \bottomrule
    \end{tabular}
\end{table}

\subsection{ How well can CHIME fix ChatGPT inaccuracies? (RQ3)}
Table \ref{tab:CHIMEresult} presents the assessments of the correctness of CHIME in our benchmark by offering overall results, across the three types of queries and also across the five task types in our benchmark. Overall, CHIME offers around 30.3\% improvement over the RAG-based ChatGPT pipeline from Section \ref{sec:rag}. The improvement is consistent across all five task types, with issue analytics and backlog tasks benefiting the most from CHIME. CHIME showcases enhancements over ChatGPT across all tasks for both OpenJ9 and ElasticSearch: 29.6\% and 31.1\% improvement over ChatGPT for OpenJ9 and ElasticSearch respectively. The detailed result for each project is provided in our online appendix. 

We manually assessed the responses where CHIME was inaccurate and observed three main reasons as follows.

\noindent\textbf{Query-Directed Retrieval Failure (60.6\%)}: CHIME relies on its ability to query a database of stored data and generate SQL queries based on the provided questions. However, when user or verifying queries lack clarity, the query fails to provide clear instructions for formulating SQL queries. In such cases, CHIME produces incorrect or irrelevant responses.

\noindent\textbf{Logical Inference Errors (27\%)}: This pertains to cases where CHIME fails to accurately apply logical inference principles. It occurs when CHIME incorrectly deduces information from the data or makes faulty assumptions during reasoning. 

\noindent\textbf{Semantic Discrepancy (10.9\%)}: CHIME relies on similarity scores to match user queries with existing data or responses. However, discrepancies in semantic similarity assessments can lead to incorrect matches or associations. 

\mybox{\textbf{Summary of RQ3.} 
CHIME achieves an average correctness of 66.7\% and an improvement of 30.3\% over a RAG-based ChatGPT on our benchmark of bug report questions.
}


\subsection{How do individual components in CHIME perform? (RQ4)}
We ran each component of CHIME individually and determined the contribution of the component within the pipeline. In Table \ref{tab:CHIMEComponentresult}, we show the performance of each component per query type and also show whether the component offered an improvement over a RAG-based ChatGPT. We discuss how we ran each component while analyzing the results below.

\begin{table}[!t]
    \centering
    \caption{Correctness of components of CHIME. Column `Improv' shows percent improvement over RAG-based ChatGPT}
    \label{tab:CHIMEComponentresult}
    \begin{tabular}{lccccc}
        \toprule
        \textbf{Component} & \textbf{Y/N} & \textbf{Fact} & \textbf{Sum} & \textbf{Total} & \textbf{Improv} \\ 
        \midrule
        Issue Preprocessor & 59.0\% & 43.9\% & 46.4\% & 48.1\% & +11.7\% \\
        Query Preprocessor & 57.0\% & 35.1\% & 45.2\% & 42.5\% & +6.1\% \\
        Response Validator & 55.0\% & 36.0\% & 47.6\% & 43.0\% & +6.6\% \\
        \midrule
        CoVe & 58.0\% & 28.9\% & 38.1\% & 37.9\% & +1.5\% \\
        MT & 69.0\% & 37.7\% & 53.6\% & 48.5\% & +12.1\% \\
        \bottomrule
    \end{tabular}
\end{table}

\noindent\textbf{Issue Preprocessor.} In our CHIME pipeline, we kept this component and removed the other two components (i.e., Query Processor and Response Validator). Hence, issue reports are preprocessed by this component and then stored in the database. From here, we utilize RAG-based ChatGPT for Q\&A. The integration of the Issue Processor enhances the accuracy of the baseline GPT model by 11.7\% on average for both projects. This improvement is particularly notable in technical question comprehension and analysis tasks.

\noindent\textbf{Query Preprocessor.} Similar to the above setup, we only kept this component and removed the other two components in our CHIME pipeline (i.e., Issue Preprocessor and Response Validator).  On average, this process demonstrates an improvement of 6.1\% over a RAG-based ChatGPT. 

\noindent\textbf{Response Validator.} We used it to validate responses from an RAG-based ChatGPT. Overall, this component contributed to a 6.6\% improvement over a RAG-based ChatGPT. The bottom two rows in Table \ref{tab:CHIMEComponentresult} further illustrate the performance of the two modules in the Response Validator, i.e., CoVe and MT. Interestingly, MT as an individual module worked even better than the Response Validator component. MT offered a 12.1\% improvement over RAG-ChatGPT while CoVe offered a 1.5\% improvement. However, we kept the combinations of CoVe and MT in the response validator, because CoVe + MT may become more useful for other repos where responses may need a sequence of challenges via both CoVe and MT. For example, when the responses from CoVe contain references to the fact (but with incorrect summarization), MT can double-check those facts via follow-up mutated questions. 


As we can see from Tables \ref{tab:CHIMEresult} and \ref{tab:CHIMEComponentresult}, CHIME as an end-to-end pipeline offers 30.3\% improvement over RAG-based ChatGPT, while none of the individual components in CHIME could offer more than 12\% improvement over RAG-based ChatGPT. This means that the ensemble of all the components in CHIME's pipeline helped the fixing of one's mistake by others. For instance, when a user queries ``List all pending issues'' rather than simply providing the count of pending issues, the transformed query from Query Preprocessor prompts the system to generate a list of issue numbers, which increases its accuracy. 

\mybox{\textbf{Summary of RQ4.} 
Each component in CHIME can offer an improvement over a RAG-based ChatGPT by correcting the inaccuracies in ChatGPT responses. The components work best when they are all put together in CHIME as an end-to-end pipeline.
}

\subsection{Would responses from CHIME favored like those from ChatGPT when both are correct? (RQ5)} \label{sec:effectsurvey}
A comparative study was conducted to assess the practical efficacy of CHIME, involving 31 participants. The majority (93\%) had 0-5 years of experience in the software industry, with 57\% being software developers and 33\% researchers.
\subsubsection{Survey Setup}

Participants were presented with two random questions from each task in our benchmark dataset. The questions include a summarization of the failure of an issue, similarities between multiple issues, recurring themes in a component, pending issues, identification of error-prone components, unresolved or blocker issues, guidelines for labeling, and label suggestions. For these questions, responses from both CHIME and ChatGPT, along with links to associated bug reports, were provided for evaluation. To ensure a fair comparison, only questions with correct responses from both systems were selected. Participants were then asked to rate the correctness and perceived usefulness of the responses in addressing software bug-related queries. The survey questions are provided in our online appendix.

\subsubsection{Survey Result}
Participant feedback in Table \ref{tab:usefulness} indicates that CHIME was the preferred choice for the majority of tasks when the answers were correct and selected more frequently for 6 out of 10 questions. It was favored in issue analysis (T1) with a 79\% participant preference. On average, for this task, 63.6\% of participants found the responses to be comprehensive and covering all necessary aspects, 45.1\% felt that they provided additional information helpful for a better understanding of the problem, and 33.3\% thought the responses were clear and easy to follow. Their preference also extended to issue summarization (T3) with a 65\% preference and to issue labeling (T4), with a 63\% preference. However, for the issue trending task (T2), there was a slight preference for ChatGPT. Nonetheless, for two questions, participants seemed undecided, indicating a comparable level of usefulness between CHIME and ChatGPT when the responses were correct.

\begin{table}[!t]
\centering
\caption{Selection Preference of CHIME and ChatGPT Provided Correct Responses across Tasks }

\renewcommand{\arraystretch}{1.3} 
\begin{tabular}{l l}
\toprule
\textbf{T\#} & \color{gray!30}\rule{4pt}{6pt} \color{black}  \textbf{ChatGPT} \color{gray!60}\rule{4pt}{6pt} \color{black} \textbf{CHIME} \\ 
\midrule 
Overall & \custombarchartwithvaluetwobar{40}{60} \\
\midrule
T1 - Issue Analytics & \custombarchartwithvaluetwobar{21}{79} \\
T2 - Issue Trend & \custombarchartwithvaluetwobar{58}{42}\\ 
T3 - Issue Summary &  \custombarchartwithvaluetwobar{35}{65}\\
T4 - Issue Labeling & \custombarchartwithvaluetwobar{37}{63}\\
T5 - Issue Backlog & \custombarchartwithvaluetwobar{48}{52}\\
\bottomrule
\end{tabular}

\label{tab:usefulness}
\end{table}

\mybox{\textbf{Summary of RQ5.} 
In a comparative study with 31 participants, CHIME responses were preferred over a stand-alone ChatGPT for the majority of tasks when both provided correct answers. This preference was particularly evident for tasks related to issue analysis, summarization, and labeling.

}

\section{Discussion}\label{sec:discussion}

\subsection{Accuracy of our CFG} \label{sec:accucfg}
We evaluated the CFG-based stack trace parsing by assessing the 80 issue reports that we used to create our benchmark dataset. The CFG is designed to identify key elements, such as exception types, messages, and code details (e.g., class/method/file names, etc.). For each stack trace, we checked whether the parser found all key elements as expected. We used three metrics to compute accuracy: precision, recall, and F1-score. Precision is the ratio of correctly identified elements to the total elements identified by the parser. Recall is the ratio of correctly identified elements to the total actual elements in the stack trace. F1-score ($F1$) is the harmonic mean of precision and recall. We manually created a list of the expected elements for accurate comparison for each of the 80 issue reports. We observed an average precision of 0.99 and recall of 0.91 (F1-score = 0.93).  The few errors in parsing were mainly due to the limitations in our regular expressions used in the CFG parser, and the variations in stack trace formats across issue reports. Our replication package contains the details of the assessment.

\subsection{Threats to Validity}\label{sec:threats}
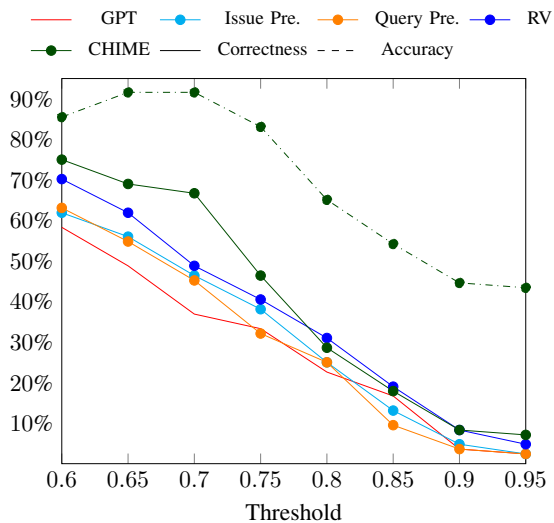
\begin{figure}[!t]
    \centering
    \begin{tikzpicture}[scale=0.9]
    \begin{axis}[
        xlabel={Threshold},
        xmin=0.6, xmax=0.95,
        ymin=0, ymax=95,
        xtick={0.6,0.65,0.7,0.75,0.8,0.85,0.9,0.95},
        ytick={10,20,30,40,50,60,70,80,90},
        ytick style={opacity=0},
        yticklabel={\pgfmathprintnumber{\tick}\%},
        legend style={
        at={(0.5,1.2)},
        anchor=north,
        legend columns=4,
        font=\footnotesize,
        draw=none,
        row sep=0.2em,
        column sep=0.5em,
        /tikz/every even column/.append style={column sep=0.2em}, 
    },
        grid style=dashed,
    ]
    
    \addplot[color=red] coordinates {
        (0.6,58.3)(0.65,48.8)(0.7,36.9)(0.75,33.3)(0.8,22.6)(0.85,16.7)(0.9,3.6)(0.95,2.4)
    };
    \addlegendentry{GPT}
    
    \addplot[color=cyan, mark=*] coordinates {
       (0.6,61.9)(0.65,56.0)(0.7,46.4)(0.75,38.1)(0.8,25.0)(0.85,13.1)(0.9,4.8)(0.95,2.4)
    };
    \addlegendentry{Issue Pre.}
    
    \addplot[color=orange, mark=*] coordinates {
         (0.6,63.1)(0.65,54.8)(0.7,45.2)(0.75,32.1)(0.8,25.0)(0.85,9.5)(0.9,3.6)(0.95,2.4)
    };
    \addlegendentry{Query Pre.}
    \addplot[color=blue, mark=*] coordinates {
        (0.6,70.2)(0.65,61.9)(0.7,48.8)(0.75,40.5)(0.8,31.0)(0.85,19.0)(0.9,8.3)(0.95,4.8)
    };
    \addlegendentry{RV}
    
    \addplot[color=green!30!black, mark=*] coordinates {
        (0.6,75.0)(0.65,69.0)(0.7,66.7)(0.75,46.4)(0.8,28.6)(0.85,17.9)(0.9,8.3)(0.95,7.1)
    };
    \addlegendentry{CHIME}

    \addlegendimage{color=black,solid}
    \addlegendentry{Correctness}
    \addlegendimage{color=black, dashed}
    \addlegendentry{Accuracy}
    
   \addplot[color=green!30!black, dashdotted, mark=*]
        coordinates {
        (0.6,85.5)(0.65,91.6)(0.7,91.6)(0.75,83.1)(0.8,65.1)(0.85,54.2)(0.9,44.6)(0.95,43.4)        
        };

    \end{axis}
    \end{tikzpicture}
    \caption{Impact of Threshold on Similarity Analysis. Here, Issue Pre. = Issue Preprocessor, Query Pre. = Query Preprocessor, RV = Response validator }
    \label{fig:threshold_analysis}
\end{figure}
Concerns regarding \textbf{construct validity} arise from the benchmark's design. However, we derived our benchmark queries from survey responses. The selection of issues from OpenJ9 and ElasticSearch may affect the generalizability of the findings across various software engineering contexts. The participant pool in surveys might not comprehensively represent the diverse perspectives in the broader software engineering community. Finally, the methodology used for evaluating CHIME could affect the accuracy and objectivity of our effectiveness assessment and may introduce \textbf{methodological bias}. However, we have analyzed the accuracy rates of summary queries across various similarity threshold values of CHIME, ranging from 0.60 to 0.95, and chose the threshold of 0.7 as it gives the highest accuracy (see Figure \ref{fig:threshold_analysis}), an optimal trade-off between capturing relevant information and minimizing false positives.  


\section{Conclusion}\label{sec:conclusion}
We have introduced CHIME to mitigate the inaccuracy of  ChatGPT response during bug report exploration. CHIME demonstrates 30.3\% improvements over ChatGPT in terms of providing more correct responses for bug exploration tasks. {Our industrial partner is working on deploying CHIME as a Slack bot. The conceptualization of CHIME originated from an internal demo of a similar chatbot created by our partner one year ago. Their initial chatbot lacked the required accuracy, which we sought to address by developing CHIME. To further motivate the need for such a chatbot beyond our industrial partner, we conducted a survey of 47 software practitioners (see Section \ref{sec:chatbotneed}). The survey findings highlight the necessity of such chatbots in the real world. Feedback from 31 industry participants, presented in Section \ref{sec:effectsurvey} shows that CHIME is preferred for its ability to analyze, summarise, and label issues. Our industrial partner was involved in the design and evaluation of CHIME, which was crucial for advancing CHIME from the proof-of-concept stage to the current deployment stage within the company. Like any innovation, we expect to improve CHIME in an agile manner, i.e., based on user feedback once deployed.

In the future, we will also focus on expanding CHIME's grasp of more technical terminologies and other documents. To handle other documents, in CHIME we will improve the issue preprocessor module e.g., to separate code and textual contents, and to adapt the CFG to handle code snippets/traces/crash dumps in those documents or using a static partial program analyzer to handle code examples in API documentation. We expect that the other modules in CHIME can be used with minimal changes.

\section*{Data Availability} The code and data used for this study can be found here: \url{https://bit.ly/4fyaMIP}.

\bibliographystyle{IEEEtran}
\bibliography{References}

\end{document}